\documentclass[a4paper,11pt]{article}

\usepackage{pos}
\usepackage{enumitem}
\usepackage{graphicx}
\usepackage{subcaption}
\usepackage{multicol}
\usepackage{float}

\usepackage{etoolbox} 
\patchcmd{\thebibliography}
  {\labelsep}
  {\itemsep 0pt \labelsep}
  {}{}

\usepackage{lipsum} 

\title{Performance Study of the IceCube Upgrade Camera System}

\ShortTitle{Performance Study of the IceCube Upgrade Camera System}

\author{The IceCube Collaboration \\{\normalsize \normalfont(a complete list of authors can be found at the end of the proceedings)}\\}

\emailAdd{rott@physics.utah.edu}
\emailAdd{pmj0324@g.skku.edu}
\emailAdd{karl.jansson@uclouvain.be}
\emailAdd{seo.won.choi@utah.edu}

\abstract{
The IceCube Upgrade Camera System is a novel calibration system designed to calibrate the IceCube detector by measuring the optical properties of the Antarctic ice. The system comprises nearly 2,000 cameras and illumination LEDs, which are present on every D-Egg and mDOM—the newly designed optical modules for the IceCube Upgrade. These units, deployed across the IceCube Upgrade volume, will capture transmission and reflection images that can be used to characterize the optical properties of both the refrozen ice within drill holes and the bulk ice between strings. Additionally, the images can aid in determining the positions of the optical modules the camera systems are mounted on. To maximize the system’s performance, various image analysis methodologies have been explored, ranging from classical maximum likelihood estimation to AI-based approaches using neural networks. In this study, we present preliminary results on the performance of these methods based on images generated by a simulation tool developed specifically for this system.

\vspace{4mm}

{\bfseries Corresponding authors:}
Carsten Rott$^{1*}$, 
Minje Park$^{2}$, 
Matti Jansson$^{3}$, 
Garrett Iverson$^{1}$, 
Seowon Choi$^{1}$\\
{$^{1}$ \itshape Department of Physics and Astronomy, University of Utah, Salt Lake City, UT 84112, USA}\\
{$^{2}$ \itshape Department of Physics, Sungkyunkwan University, Suwon 16419, Republic of Korea}\\
{$^{3}$ \itshape CP3, Université catholique de Louvain, Louvain-la-Neuve, Belgium}\\
$^*$ Presenter
}

\ConferenceLogo{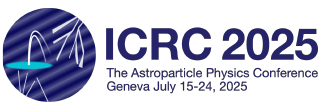}

\FullConference{39th International Cosmic Ray Conference (ICRC2025)\\
 15–24 July 2025\\
Geneva, Switzerland\\}

\begin{document}
\raggedbottom
\maketitle

\section{Introduction}
\label{sec1}
The IceCube Neutrino Observatory~\cite{DetectorPaper}, located at the geographic South Pole, is a cubic-kilometer-scale detector instrumented with 5,160 digital optical modules (DOMs) arranged on 86 vertical strings between depths of $1.5~\mathrm{km}$ and $2.5~\mathrm{km}$ in the Antarctic ice. The IceCube Upgrade~\cite{ICRC2019:ICU-project} will consist of seven additional strings, numbered 87 to 93, at the center of the existing array. These strings will carry newly designed optical modules, arranged in a denser configuration. While one of the primary goals of the Upgrade is to enhance sensitivity to lower-energy neutrinos, another major objective is to calibrate the IceCube detector. 
In particular, limited knowledge about the optical properties of the Antarctic ice, which serves as a Cherenkov detection medium, is the source of uncertainty in neutrino event reconstruction~\cite{ICRC2019:ICU-project}. To better constrain these uncertainties, the IceCube Upgrade has added new calibration systems to complement and extend the well-established flasher-based calibration.

The IceCube Upgrade Camera System~\cite{ICRC2019:ICU_camera} is a set of such devices composed of wide field-of-view cameras and illumination LEDs. Each of the newly designed optical modules, the D-Eggs~\cite{ICRC2021:D-Egg} and mDOMs~\cite{ICRC2021:mDOM}, is equipped with this system, resulting in nearly 2,000 cameras and LEDs deployed throughout the Upgrade volume. Depending on the combination of the camera and the LED used, the captured images aim to characterize either the bulk ice between strings or the refrozen ice within the drill hole. In addition, the images can aid in refining the positions and orientations of the optical modules they are mounted on.

\section{Upgrade Camera System}
\label{sec2}

The camera module utilizes a CMOS color image sensor manufactured by Sony (model IMX225LQR-C), which provides a full resolution of $1312\times993$. It is equipped with a fisheye lens that has an approximately $170^{\circ}$ field of view in the air. Considering the refractive index of ice and the support structure of the optical modules, their expected field of view after being deployed in the ice is $120^{\circ}$ and $90^{\circ}$ for D-Egg and mDOM cameras, respectively. The illumination module employs a $451~\mathrm{nm}$ wavelength LED with a beam that has a full width at half maximum (FWHM) of approximately $90^{\circ}$. Both are engineered to operate at temperatures as low as $-40^{\circ}$C. 

In the D-Egg module, three camera–LED pairs are mounted azimuthally in a horizontal ring around the lower PMT, spaced $120^{\circ}$ apart. The cameras are installed on sliding ring segments that allow for slight movement to accommodate compression of the module. After freeze-in, this compression can cause the angular separation between pairs to deviate by up to $12^{\circ}$.  In the mDOM, two camera–LED pairs are installed at 45° angles in the upper hemisphere, and one at the bottom facing downward. An additional upward-facing LED is mounted at the top. The D-Egg's camera system configuration supports inter-string imaging, using LEDs and cameras on different strings. While the mDOM's design enables imaging within individual drill holes, using a downward-facing camera with either its adjacent LED or the top LED of the mDOM positioned below.

\begin{figure}[H]
    \centering
    \includegraphics[width=0.8\linewidth]{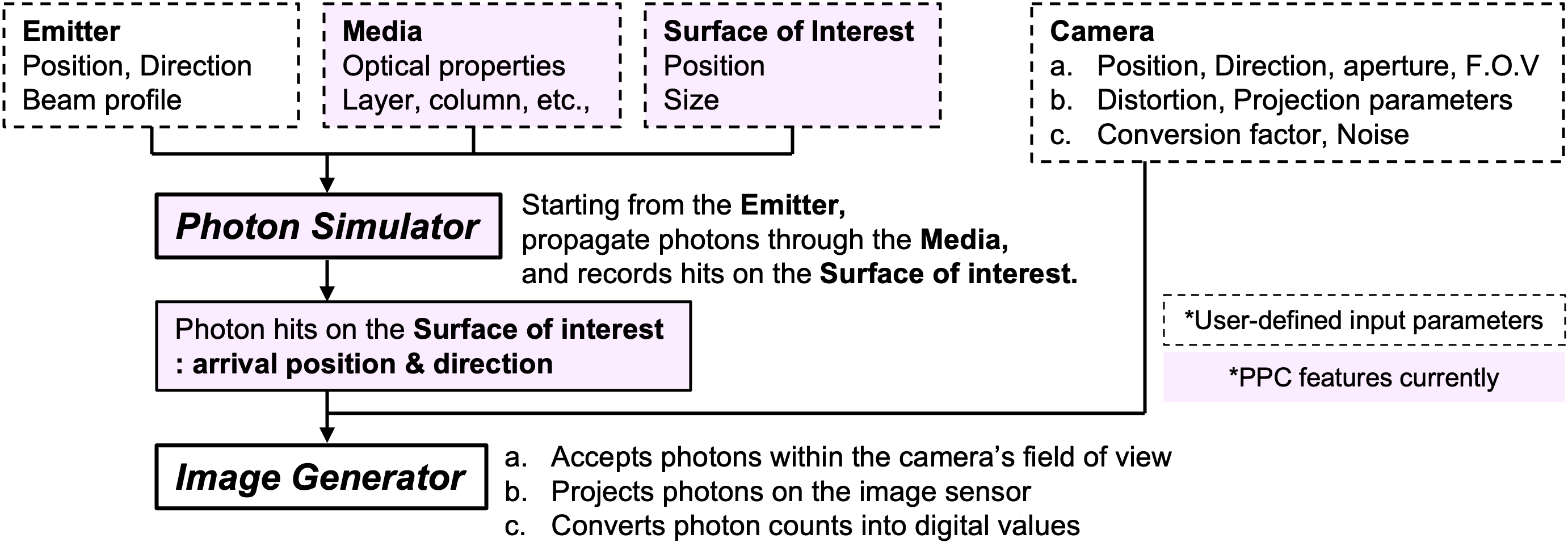}  
    \caption{\raggedright CamSim simulates photon propagation using user-defined configurations. The resulting photon hits are projected onto the image sensor.  Parameters (a), (b), and (c) under the \textbf{Camera} block  correspond to steps (a), (b), and (c) in the \textbf{Image Generator}.Pink-shaded boxes indicate components currently implemented in the PPC photon simulator.}
    \label{fig:camsim_flow}
\end{figure}

To develop an analysis framework for measuring ice properties from camera images, we built a dedicated simulation framework, CamSim~\cite{ICRC2023:CamSim}. (See Figure~\ref{fig:camsim_flow}) CamSim generates synthetic images based on user-defined light sources, media properties, and camera models. The simulation starts by defining three components: the emitter, medium, and surface of interest.

Emitter parameters include the light source’s position, orientation, and angular intensity profile. The medium is described by optical properties such as scattering and absorption lengths; recent updates allow layered and spatially varying media, including inhomogeneous structures. The surface of interest is a virtual detection plane that records photon arrival positions and, in this study, corresponds to the surface of the optical module.

Once defined, CamSim uses the Photon Propagation Code (PPC)~\cite{ppc} to simulate photon transport through the medium. Scattering and absorption are applied according to medium parameters, and arrival positions and angles of photons reaching the surface are recorded as \textit{photon hits}.

These hits are then projected onto the image sensor. CamSim filters photons based on the camera’s position, orientation, and field of view, selecting only those entering the aperture. The selected photons are projected using defined parameters, including lens distortion, and converted to digital pixel values via a conversion factor that accounts for the sensor’s photon-to-electron response. A noise model based on lab measurements at \(-40^\circ\mathrm{C}\) adds Poissonian fluctuations based on pixel intensity, enabling both realistic image generation and statistical sampling from a single simulation.

An important design choice in CamSim is to apply the camera model only after photon propagation. This modular structure allows a single, computationally expensive photon simulation to be reused to generate images for multiple camera models or viewpoints. It can also be used to generate multiple noise realizations from the same photon hit map, enhancing statistical robustness in downstream analysis. While CamSim currently relies on PPC for photon propagation and surface interaction handling, future updates aim to extend compatibility to other propagation engines using the same workflow.

\section{Ice Property Measurement}
\label{sec4}
The primary task of the IceCube Upgrade Camera system is to measure the properties of the detector medium, Antarctic ice, thereby reducing the uncertainty in neutrino event reconstruction. For this purpose, we developed a deep learning model, \textbf{OPTICUS}-Bulk \& Hole, that can infer the desired optical properties of ice from images.

\subsection{Deep Learning Model -- OPTICUS}
\label{sec4_1}
Geometry estimation relies on clear visual cues like the relative direction of the LED, whereas optical property estimation involves complex, non-linear patterns that are difficult to describe with hand-crafted features. Deep learning models are well-suited for this task. Convolutional Neural Networks (CNNs) extract subtle, spatially distributed features, while transformer-based architectures like Vision Transformer(ViT)~\cite{vit_art}) capture global contextual relationships across the image. 

\begin{figure}[ht!]
    \centering
    \includegraphics[width=0.8\linewidth]{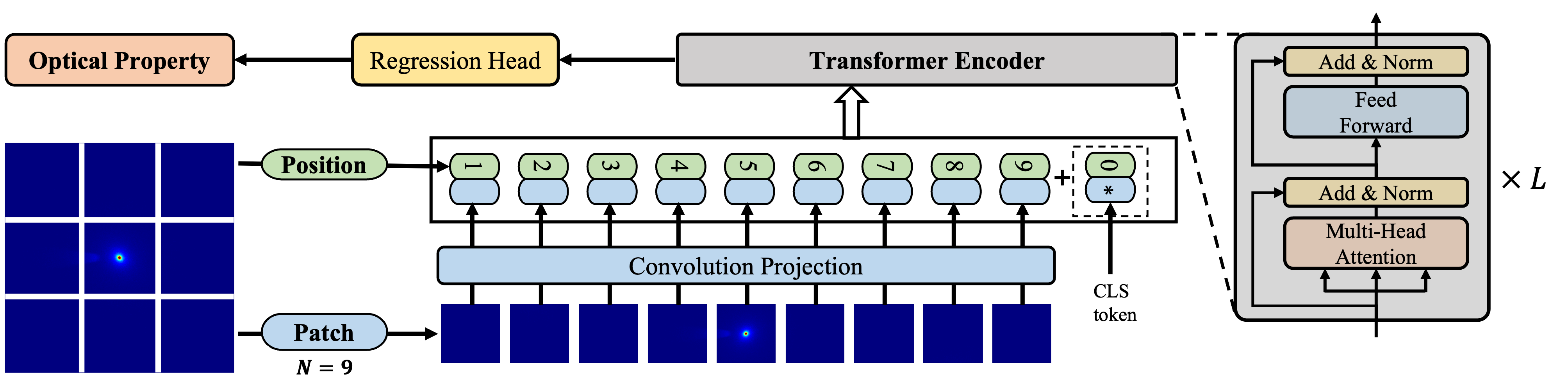}
    \caption{Overall structure of OPTICUS, inspired by the Vision Transformer (ViT) \cite{vit_art}. The input image is split into patches ($N=9$ shown as an example) and embedded via a convolutional projection. With positional embeddings, the resulting sequence of embedded vectors including CLS token is processed by a stack of $L$ encoder layers . Finally, the CLS token from the last encoder layer is passed through a regression head to predict the optical property.}   
    \label{fig:opticus_str}
\end{figure}

To leverage both the local feature extraction of CNNs and the global attention of Vision Transformers, we developed a hybrid model named \textbf{OPTICUS} (\emph{Optical Property Transformer for IceCube Upgrade Camera System}). As illustrated in Figure~\ref{fig:opticus_str}, the model first partitions the input image into $N$ patches and projects each into an embedding vector via a convolution projection. Next, a learnable CLS token is appended to the sequence, and for all $N+1$ tokens ($N$ embedding vectors + class token) positional embeddings are applied to preserve spatial context. These tokens are processed by a stack of $L$ Transformer encoder layers~\cite{vaswani2017attention}.  Each layer (depicted in the right inset of Figure \ref{fig:opticus_str}) consists of a multi-head self-attention mechanism followed by a position-wise feed-forward network. Both sublayers are equipped with residual connections and layer normalization. 
Finally, the CLS token, which serves as a global representation of the input image, is then fed into a regression head that outputs a single value representing the optical property.

\subsection{Bulk Ice Model}
\label{sec4_2}

To develop a model for estimating the bulk ice scattering length, we generated simulated images using a configuration in which a camera from a D-Egg captures light emitted horizontally by an LED from another D-Egg. The two modules are placed 29.8\,m apart at the same depth, approximating the inter-string spacing of the IceCube Upgrade. With fixed positions, the camera was rotated in steps of $20^\circ$ across a range of $\pm60^\circ$. For each geometric configuration, $10^{12}$ photons of 470\,nm wavelength were injected, while the bulk ice scattering length was varied from 20\,m to 100\,m in 1\,m increments. Based on these 567 simulations, 100 noise realizations were applied to each image, resulting in a total of 56{,}700 images. Based on this dataset, the \textbf{OPTICUS}-Bulk model was configured with \( N = 100 \) and \( L = 5 \).

\begin{figure}[h]
     \centering
    \includegraphics[width=0.85\linewidth]{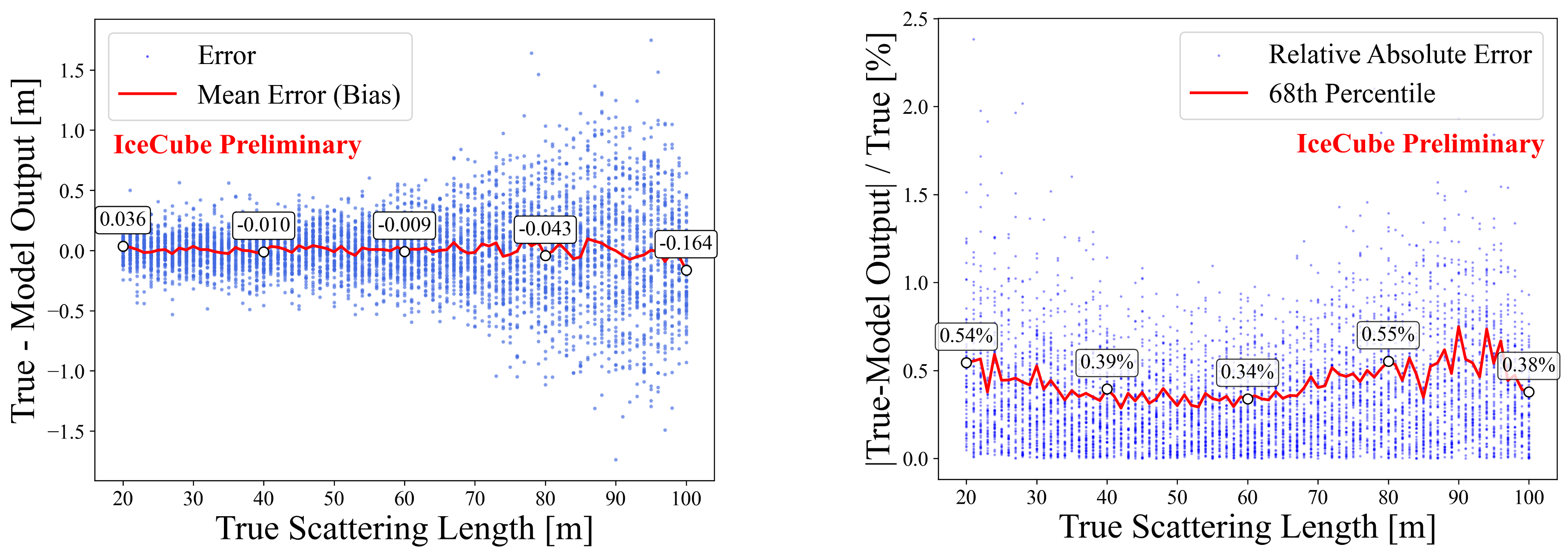}
        \caption{Prediction performance of the \textbf{OPTICUS}-Bulk on simulated test data. Prediction bias (left) was assessed using the error defined as $\text{true} - \text{model output}$\,[m]. For each true scattering length, the bias (mean error) was computed, and a few examples are highlighted with circles and text for visual guidance. Prediction accuracy (right) was assessed using the absolute relative error $|\text{true}-\text{model output}|/\text{true}[\%]$. For each true scattering length, 68-percentile values are calculated and highlighted with white circles and text. Optical properties other than the bulk ice scattering length are not included.}
        \label{fig:bulk_result}
\end{figure}

Each image was labeled solely with the bulk ice scattering length, and 90\% of the dataset was used for training. The prediction performance of the trained model, \textbf{OPTICUS}-Bulk, was evaluated on the remaining 10\%. As shown in Figure~\ref{fig:bulk_result}, the model accurately predicts scattering lengths across the full simulated range. For true scattering lengths of $20~\mathrm{m}-100~\mathrm{m}$, the average prediction errors are tightly centered around zero, indicating minimal bias. The accuracy is demonstrated with relative error and remains below 0.5\% for most cases. The overall 68\% containment across the entire range is 0.42\%.

\subsection{Hole Ice Model}
\label{sec4_3}

The ``hole ice'' refers to the refrozen column of ice within each IceCube borehole. A region near its center, known as the “bubble column,” exhibits significantly shorter scattering lengths than the surrounding ice. This column is thought to form primarily from residual air bubbles released during the freezing process, as the hole solidifies inward from the borehole walls.

To develop a model for estimating the bubble column scattering length, we generated simulated images using a configuration in which the downward-facing camera of an upper mDOM observes light emitted by the upward-facing LED of a lower mDOM. Both mDOMs were positioned at the center of the hole ice. The bubble column was modeled at a distance of \(10\,\mathrm{cm}\) from the center of the hole, with a radius equal to half that of the mDOM. The scattering length of the bubble column was varied from \(2\,\mathrm{cm}\) to \(30\,\mathrm{cm}\) in \(2\,\mathrm{cm}\) steps, resulting in 15 distinct configurations. For each case, five independent simulations were performed to generate base images, and noise was added 200 times per image, yielding a dataset of 15{,}000 images in total. Based on this dataset, the \textbf{OPTICUS}-Hole model was configured with \( N = 100 \) and \( L = 3 \).

\begin{figure}[h]
     \centering
         \includegraphics[width=0.85\linewidth]{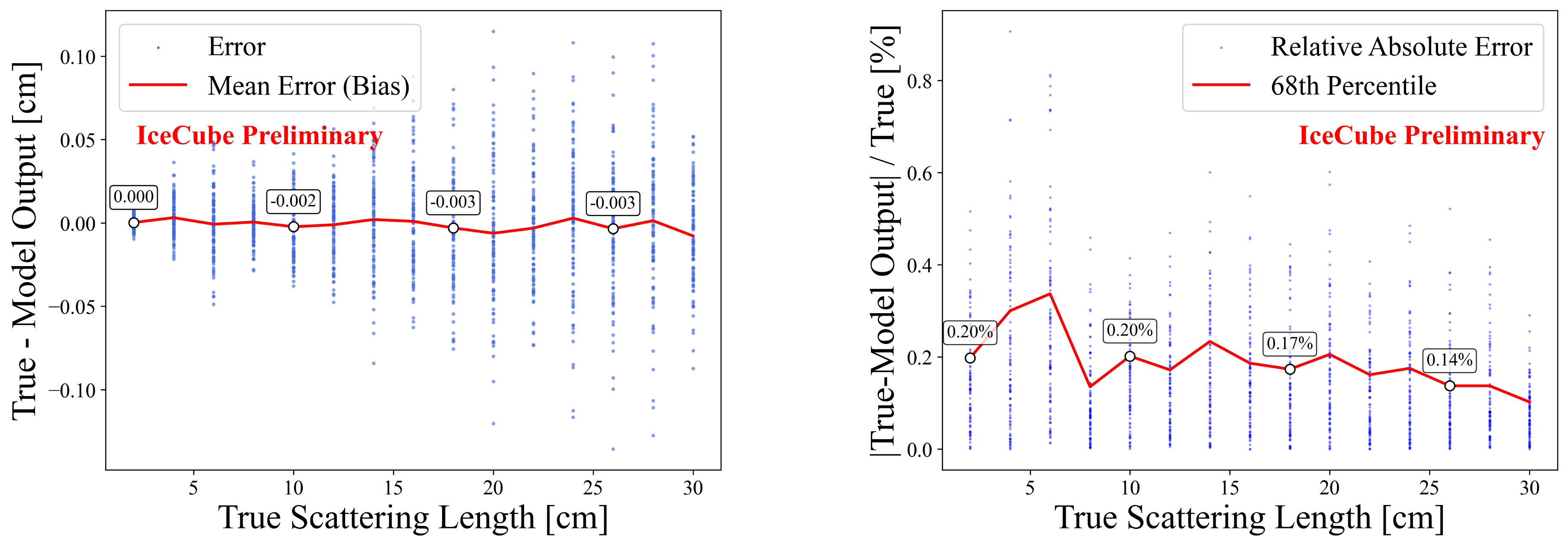}
        \caption{Prediction performance of \textbf{OPTICUS}-Hole on simulated test data. Prediction bias (left) and accuracy (right) were assessed in the same manner as for \textbf{OPTICUS}-Bulk. For each true scattering length, the 68th percentile of the error distribution was computed, with several examples highlighted with text for visual clarity. Optical properties other than the bubble column scattering length are not included.}
        \label{fig:hole_result}
\end{figure}

The prediction performance of the trained model, \textbf{OPTICUS}-Hole, is demonstrated in Figure~\ref{fig:hole_result}. Again, bias is assessed using prediction errors, as shown in the left plot. The errors are tightly centered around zero, indicating that there is no notable bias across the 2--30\,cm range. The absolute relative error distribution, shown on the right plot, illustrates the prediction accuracy. The 68\% percentile of each bin remains below 2\% in all cases, and below 1\% across most of the range. The overall 68\% containment across the entire range is 0.18\%.

Both \textbf{OPTICUS}-Bulk and \textbf{OPTICUS}-Hole demonstrate percentage-level prediction accuracy. Considering that the simulation was based on a single camera-LED pair, this level of performance suggests strong potential for even greater accuracy when applied to real measurements, where multiple camera-LED combinations will provide redundant and complementary views. Nevertheless, it is important to note that the reported performance reflects statistical uncertainties only, as the validation data was generated under the same simulation configuration as the training data. Other limitations and plans to address them will be discussed in the following section.
\section{Geometry Calibration}
\label{sec3}

The positions and directions of the deployed modules affect signal reconstruction as well as calibration measurements using the cameras. Positions  will be measured using flashers \cite{Abbasi:20231J} or where available acoustic modules \cite{Abbasi:2021Ku}, with orientations measured using in-module accelerometers (tilt) and magnetometers (azimuthal)\cite{gen2TDRpart2}. Using images that are recorded to measure bulk ice properties, we can get an independent measurement of orientation and using Bayesian inference we can get a refinement of the geometry measurement. To do this we use pictures taken by cameras in 7 D-Eggs at the same depth. As there is a reliable method to obtain the relative depth \cite{Aartsen:2016nxy}, we only try to constrain the modules' positions in the plane. With the analysis running in layers with one module from each string, we will identify the DOMs with the string numbers. 

First, the LED pixels are extracted from camera images. This is done by calculating the center of mass of the pixels with a gradient magnitude within 10\% of the highest gradient magnitude in the image. From the LED pixel an angle can be calculated using calibrated projection parameters. 
Once the angle to two neighboring modules is measured, we add them to get an angle, repeating for all triples of DOMs we get a set of up to 18 angles, which is referred to as the
observed angles $\theta_{\mathrm{obs}}$.

For each possible configuration of DOM positions, the corresponding angles $\theta_{\mathrm{cal}}$ can be calculated. Given $\theta_{\mathrm{cal}}$ and the angle uncertainty $\sigma_\theta$, which is estimated from simulation to be \(1^\circ\), the likelihood of the observed angles $\theta_{\mathrm{obs}}$ is given in Equation~\ref{eq:lik}, and the posterior in Equation~\ref{eq:post}.

\begin{equation}
    L(\theta_{\mathrm{obs}}|x) = \prod_{i=1}^{18} \frac{1}{\sigma_\theta\sqrt{2\pi }}\mathrm{exp}\big(-\frac{(\theta_{\mathrm{obs},i}-\theta_{\mathrm{cal},i})^2}{2\sigma_\theta^2}\big)
    \label{eq:lik}
\end{equation}

\begin{equation}
    P(x|\theta_{\mathrm{obs}}) = \frac{L(\theta_{\mathrm{obs}}|x)Prior(x)}{\int dx \; L(\theta_{\mathrm{obs}}|x) Prior(x)}.
    \label{eq:post}
\end{equation}

 But with 7 modules in each layer we have 14 dimensions, so a naive numerical integration will not work. We here use Monte Carlo integration, where each dimension has been sampled using a Gaussian.  

\begin{figure}[h]
    \centering
    \includegraphics[width=0.8\textwidth]{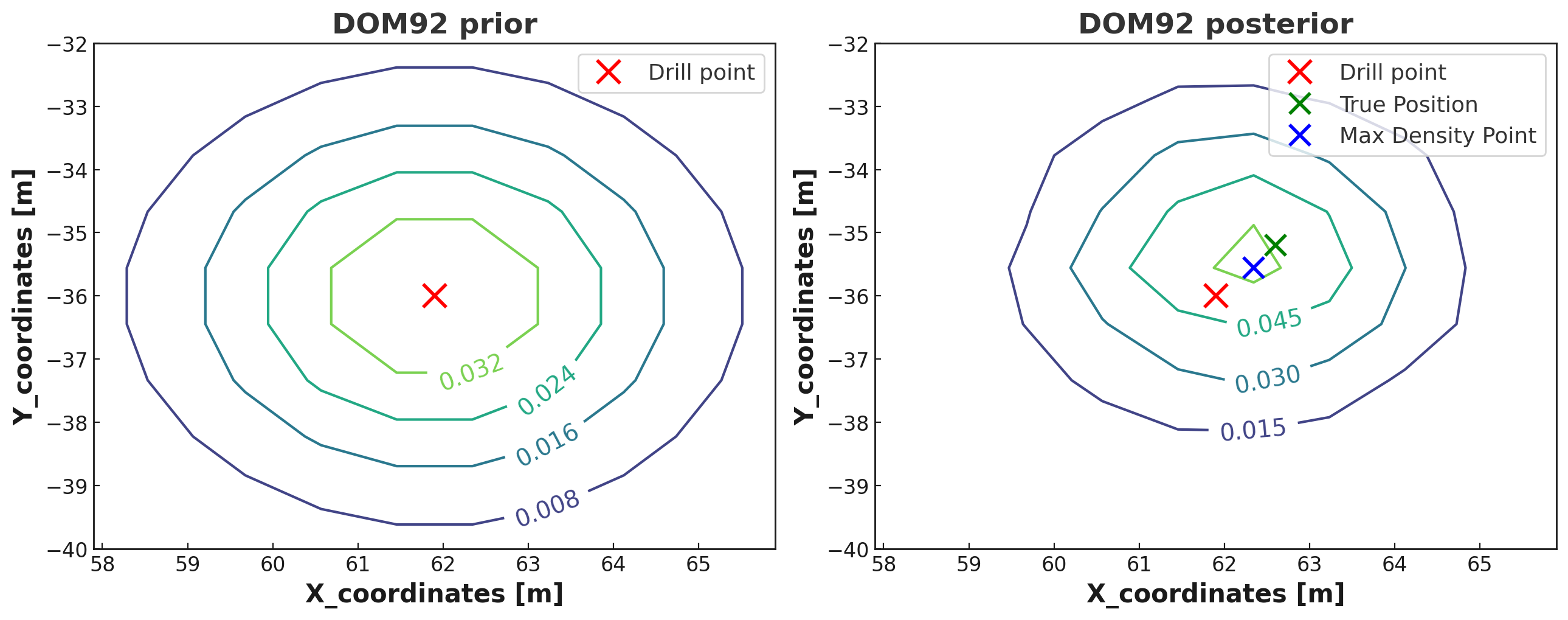}
    \caption{Contour plots of the prior (left) and posterior (right) probability distributions for DOM 92. The red cross indicates the drill point, the green cross the true position, and the blue cross the posterior best fit. The contours shrink and move closer to the true position, though some uncertainty remains due to Monte Carlo integration used in marginalization.}
    \label{fig:baysian_result}
\end{figure}

To visualize the results, we marginalize all but the two coordinates of a module. Figure~\ref{fig:baysian_result} shows the result from simulated data. The method can be applied with any number of measured angles and is robust to missing data, but for this test, only the cameras and LEDs of three modules were simulated. Table~\ref{tab:baysian_result} presents the positional errors from the prior and posterior best-fit estimates. The posterior shows a clear improvement, with reduced distance errors indicating higher positional accuracy.

\begin{table}[h]
    \centering
    \resizebox{1.0\textwidth}{!}{%
    \renewcommand{\arraystretch}{1.0} 
    \begin{tabular}{|c|c|c|c|c|c|}
        \hline
        \textbf{DOM} & \textbf{True Point (x, y)} & \textbf{Drill Point (x, y)} & \textbf{Prior Distance (m)} & \textbf{Best Fit Point (x, y)} & \textbf{Posterior Distance (m)} \\
        \hline
        \textbf{88} & (47.3, -57.0) & (48.3, -56.5) & 1.12 & (47.0, -57.9) & 0.95 \\
        \textbf{92} & (62.6, -35.2) & (61.9, -36.0) & 1.06 & (62.4, -35.3) & 0.22 \\
        \textbf{93} & (27.0, -31.2) & (27.3, -30.5) & 0.76 & (27.0, -30.9) & 0.30 \\
        \hline
    \end{tabular}%
    }
    \caption{Comparison of initial and reconstructed positions for modules on strings 88, 92 and 93 and distances to true positions.}
    \label{tab:baysian_result}
\end{table}

\section{Conclusion and Outlook}
\label{sec5}

We evaluated the performance and sensitivity of the IceCube Upgrade Camera System for its intended role within the IceCube Upgrade. The analysis used simulated images generated by the custom CamSim framework, which models image capture under various geometric configurations and optical conditions. For geometry measurements, we showed that relative angles between DOMs, extracted from captured images, can be used within a Bayesian framework to reconstruct DOM positions with sub-meter precision. To study the ice’s optical properties, we developed Transformer-based models trained on simulated images to infer the scattering length of the bulk ice and the bubble column, respectively. The models achieved high accuracy at a sub-percentage level.

The current model, trained on simulated images with varying scattering lengths, serves as a prototype for a more comprehensive framework. We aim to expand this approach by incorporating additional optical properties into the dataset, such as absorption length, scattering phase function, and anisotropy. This will enable the development of a unified model capable of inferring multiple ice properties from a single image. While the inclusion of correlated parameters may reduce sensitivity, it will lead to a more realistic and fully integrated model. To support this effort, future simulations will integrate a refined projection mechanism that accounts for photon propagation through the full optical path, including the pressure vessel, optical gel, mechanical mount, and lens. A water tank setup is currently being constructed at the University of Utah, where a real mDOM will be imaged under controlled conditions using a movable light source on an x–y gantry system.


\bibliographystyle{ICRC}
\bibliography{references}

%

\clearpage

\section*{Full Author List: IceCube Collaboration}

\scriptsize
\noindent
R. Abbasi$^{16}$,
M. Ackermann$^{63}$,
J. Adams$^{17}$,
S. K. Agarwalla$^{39,\: {\rm a}}$,
J. A. Aguilar$^{10}$,
M. Ahlers$^{21}$,
J.M. Alameddine$^{22}$,
S. Ali$^{35}$,
N. M. Amin$^{43}$,
K. Andeen$^{41}$
% ...
\\
\\
% ...
$^{10}$ Universit{\'e} Libre de Bruxelles, Science Faculty CP230, B-1050 Brussels, Belgium \\
% ...
$^{16}$ Department of Physics, Loyola University Chicago, Chicago, IL 60660, USA \\
$^{17}$ Dept. of Physics and Astronomy, University of Canterbury, Private Bag 4800, Christchurch, New Zealand \\
% ...
$^{21}$ Niels Bohr Institute, University of Copenhagen, DK-2100 Copenhagen, Denmark \\
$^{22}$ Dept. of Physics, TU Dortmund University, D-44221 Dortmund, Germany \\
% ...
$^{35}$ Dept. of Physics and Astronomy, University of Kansas, Lawrence, KS 66045, USA \\
% ...
$^{39}$ Dept. of Physics and Wisconsin IceCube Particle Astrophysics Center, University of Wisconsin{\textemdash}Madison, Madison, WI 53706, USA \\
% ...
$^{41}$ Department of Physics, Marquette University, Milwaukee, WI 53201, USA \\
% ...
$^{43}$ Bartol Research Institute and Dept. of Physics and Astronomy, University of Delaware, Newark, DE 19716, USA \\
% ...
$^{63}$ Deutsches Elektronen-Synchrotron DESY, Platanenallee 6, D-15738 Zeuthen, Germany \\
$^{\rm a}$ also at Institute of Physics, Sachivalaya Marg, Sainik School Post, Bhubaneswar 751005, India \\
% ...

\subsection*{Acknowledgments}

\noindent
The authors gratefully acknowledge the support from the following agencies and institutions:
USA {\textendash} U.S. National Science Foundation-Office of Polar Programs,
U.S. National Science Foundation-Physics Division,
U.S. National Science Foundation-EPSCoR,
U.S. National Science Foundation-Office of Advanced Cyberinfrastructure,
Wisconsin Alumni Research Foundation,
Center for High Throughput Computing (CHTC) at the University of Wisconsin{\textendash}Madison,
Open Science Grid (OSG),
Partnership to Advance Throughput Computing (PATh),
Advanced Cyberinfrastructure Coordination Ecosystem: Services {\&} Support (ACCESS),
Frontera and Ranch computing project at the Texas Advanced Computing Center,
U.S. Department of Energy-National Energy Research Scientific Computing Center,
Particle astrophysics research computing center at the University of Maryland,
Institute for Cyber-Enabled Research at Michigan State University,
Astroparticle physics computational facility at Marquette University,
NVIDIA Corporation,
and Google Cloud Platform;
Belgium {\textendash} Funds for Scientific Research (FRS-FNRS and FWO),
FWO Odysseus and Big Science programmes,
and Belgian Federal Science Policy Office (Belspo);
Germany {\textendash} Bundesministerium f{\"u}r Forschung, Technologie und Raumfahrt (BMFTR),
Deutsche Forschungsgemeinschaft (DFG),
Helmholtz Alliance for Astroparticle Physics (HAP),
Initiative and Networking Fund of the Helmholtz Association,
Deutsches Elektronen Synchrotron (DESY),
and High Performance Computing cluster of the RWTH Aachen;
Sweden {\textendash} Swedish Research Council,
Swedish Polar Research Secretariat,
Swedish National Infrastructure for Computing (SNIC),
and Knut and Alice Wallenberg Foundation;
European Union {\textendash} EGI Advanced Computing for research;
Australia {\textendash} Australian Research Council;
Canada {\textendash} Natural Sciences and Engineering Research Council of Canada,
Calcul Qu{\'e}bec, Compute Ontario, Canada Foundation for Innovation, WestGrid, and Digital Research Alliance of Canada;
Denmark {\textendash} Villum Fonden, Carlsberg Foundation, and European Commission;
New Zealand {\textendash} Marsden Fund;
Japan {\textendash} Japan Society for Promotion of Science (JSPS)
and Institute for Global Prominent Research (IGPR) of Chiba University;
Korea {\textendash} National Research Foundation of Korea (NRF);
Switzerland {\textendash} Swiss National Science Foundation (SNSF).

\end{document}